\documentclass[aps,prd,final,letterpaper, twocolumn]{revtex4}

\usepackage{graphicx}                            % Graphics package
\usepackage{amsmath} 
\usepackage[hidelinks]{hyperref}

\newcommand{\bra}[1]{\left\langle #1 \right|}
\newcommand{\ket}[1]{\left|#1\right\rangle}

\newcommand{\rvec}{\mathbf{r}}
\newcommand{\ham}[1]{\mathcal{H}#1}
\newcommand{\normsq}[1]{\mathop{\lvert #1 \rvert^2}}
\DeclareMathOperator{\sinc}{sinc}

\begin{document}

%%%%%%%%%%%%%%
%%
%% Now, the document itself begins.  When you are ready to begin writing
%% your own paper, you will cut content out of the following and
%% replace it with your own.  Keep an unmodified copy of this
%% LaTeX file, though, and a hard copy of the paper by Prof.
%% Jaffe which it produces.  Comparing the template below with
%% the hard copy it produces will help you see how to handle
%% titles, sections, figures, equations, references, the bibliography
%% and much more.  Good luck with your paper. JN.
%%
%%%%%%%%%%%%%%

\title{ Maser Radiation in an Astrophysical Context}
\author{Eric~S.~Gentry}
\affiliation{ MIT Department of Physics,
 77 Massachusetts Ave.,
Cambridge, MA 02139-4307}
\date{May 6, 2013} 

\begin{abstract}
\noindent	
In this paper we will look at the phenomenon of Microwave Amplification by Stimulated Emission of Radiation (a \emph{maser} system).  We begin by deriving amplification by stimulated emission using time-dependent perturbation theory, in which the perturbation provided by external radiation.  When this perturbation is applied to an ensemble of particles exhibiting a population inversion, the result is stimulated microwave radiation. We will explore both unsaturated and saturated masers and compare their properties.  By understanding their gain, as well as the effect of \emph{line broadening}, astronomers are to identify astrophysical masers. By studying such masers, we gain new insight into poorly understood physical environments, particularly those around young and old stars, and compact stellar bodies.
			
\end{abstract}

\maketitle

\pagestyle{myheadings}
\markboth{E.S. Gentry}{ Maser Radiation in an Astrophysical Context }
\thispagestyle{empty}

\section{Introduction}

\begin{figure}
		\centering
		\includegraphics[width=\linewidth]{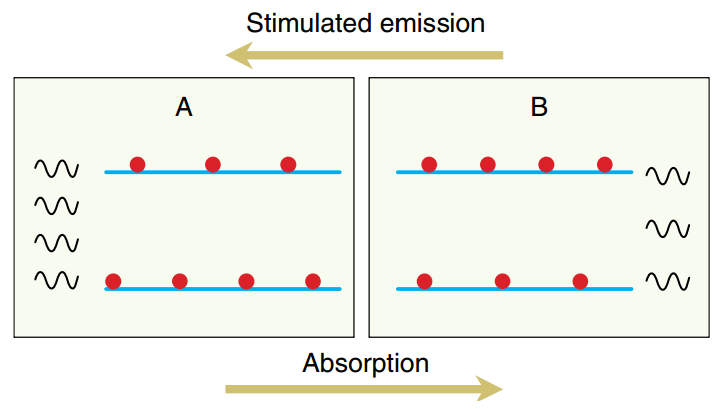}
		\caption{Absorption occurs as one photon on Side A is absorbed to raise a particle to the higher energy level.  Masers use population inversions and stimulating radiation to induce the reverse process --- an excited particle on Side B drops to a lower energy level, producing a photon in the process. Diagram produced by Elitzur \cite{ElitzurNat}.}
		\label{fig:StimulatedEmissionAbsorption_Elitzur}
\end{figure}

	Typically when electromagnetic radiation passes through a medium, the intensity is attenuated as photons are absorbed. Certain systems do the opposite however; rather than attenuating a signal they amplify it at select frequencies.  Such a phenomenon is called a \emph{laser}, Light Amplification by Stimulated Emission of Radiation (lasers in the microwave band are a special case, called \emph{masers}).  These systems are uncommon; they require energy levels to be populated in ways not possible under the conditions of thermodynamic equilibrium (they require a \emph{population inversion} of energy levels; see Figure \ref{fig:StimulatedEmissionAbsorption_Elitzur} for a pictorial explanation of population inversions and the relationship between amplification and absorption).  Something must be forcing these systems away from equilibrium.  In studying those forcing mechanisms, we benefit from the amplifying properties of lasers.  While particles in thermal equilibrium typically absorb light, systems with population inversions can become shining beacons.  For this paper we will look at their application to astrophysical systems --- systems which we have no method to directly probe, for which we must rely on observation, inference, and theory.  Astronomers typically study \emph{masers}, microwave lasers, since microwave signals tends to penetrate well through atmospheric gases on earth and interstellar dust in space \cite{Shklovsky}.
	
	In this paper, Section \ref{sec:ThermoEq} will examine why masers are uncommon physical phenomenae, and why it might be interesting to study the physical conditions to which they are sensitive.  Section \ref{sec:derivation} looks at the theory of stimulated emission, and will identify those conditions to which masers are sensitive.   That knowledge guides both our search for astrophysical masers, and our ability to infer physical conditions of distant environments. To conclude, several topics of current research will be discussed in Section \ref{sec:ResearchApplications}.

	\section{Thermodynamic Equilibrium; Why Are Masers Uncommon?}
	\label{sec:ThermoEq}
	Population inversions are necessary for amplification by stimulated emission (discussed more fully in Section \ref{sec:AstroAmp}). \emph{Population inversions} are when a higher energy level is populated with more states than a lower energy level.  The reason why this does not typically occur, and why masers are so special, is because the canonical ensemble of thermodynamics predicts energy level population ratios:
	
	\begin{equation}
		\frac{N1}{N2} = \exp \left[ - \frac{E_1 - E_2}{k_B T} \right]
	\end{equation}
	
	For a system with a ground state $\ket{1}$ of energy $E_1$ and excited state $\ket{2}$ of energy $E_2$, we expect $N_2 < N_1$. A population inverse is the exact opposite; it requires $N_2 > N_1$ which cannot be predicted by equilibrium thermodynamics.  To leave equilibrium, these systems require \emph{pumping}, the addition of states to the higher energy level faster than \emph{quenching} processes can restore equilibrium.  For gases, quenching usually comes in the form of \emph{collisional quenching} which increases with density and pressure \cite{GoldreichKwan}.
	
\section{Production of Masers}
		\label{sec:derivation}
		
		The theory underlying stimulated emission is first-order time-dependent perturbation theory.  Gases even as simple can have intrinsic electric dipoles, and if our gas particle has an intrinsic electric dipole, then an external electromagnetic field will produce a time-dependent perturbation to the system's energy.  Such a time-dependent perturbation is particularly useful, because it allows for transitions between pure energy eigenstates.  If this transition is to a lower energy level, a photon will be emitted with a frequency determined by the energy difference between the initial and final states.  In Section \ref{sec:derivation_transition_rates} we will use time-dependent perturbation theory to derive the expected transition rate, and its dependence on external factors (such as applied field strength) and intrinsic properties (such as permanent electric dipole moment of the particle).  In Section \ref{sec:AstroAmp} we look at how those transitions manifest themselves as amplification of an incident signal.

		\subsubsection{Transition Rates}
		\label{sec:derivation_transition_rates}

		For simplicity we will deal with 2 quantum states of a gas (either atomic or molecular).  We will label these orthonormal states $\ket{1}$ and $\ket{2}$ with non-degenerate energies $E_1$ and $E_2$, such that $\omega_0 \equiv \frac{ \lvert E_2 - E_1 \rvert}{\hbar} \neq 0$.  Those energies are determined by a Hamiltonian, $\ham{}$ which can include rotational, vibrational, coulombic, and spin components of the system's energy. The exact form of the Hamiltonian is not necessary (and indeed can be incredibly complex); we only need an approximate spectrum of energy levels and states.
		
		In the case of stimulated emission, a traveling electromagnetic wave produces a time-dependent perturbation, $\ham{'}$, to the Hamiltonian.  This time-dependent perturbation will typically be dominated by electric dipole interactions, rather than magnetic or higher multipole moments \cite{Siegman}. For a linearly polarized monochromatic plane wave, traveling in $+\widehat{e_x}$, we expect fields of the form:
		
		\begin{eqnarray}
			\mathbf{E} &= E_0 & \sin(\omega t - k x) \widehat{e_y}  \\
			\mathbf{B} &= \frac{E_0}{c}& \sin(\omega t - k x) \widehat{e_z}
		\end{eqnarray}
		
		Which gives an electric dipole perturbation to the Hamiltonian:
		
		\begin{equation}
			\ham{'} =  q y E_0 \sin(\omega t - k x ) 
		\end{equation}
		
		Ideally, we would like to factorize the perturbation into a time-dependent term and a space-dependent term.  This will allow us to deal with the spatial dependence separately from the periodic time dependence, which is a well-studied topic.  While $\sin(\omega t - k x )$ is not directly factorizable, the gas particle can only feel spatial variations in the electric field on the order of the particle diameter (picometers) whereas the electric field varies on much larger spatial scales (centimeters in the case of microwaves).  That difference in scales allows us to approximate the fields as being spatially independent (as far as the particles can tell):
		
		\begin{eqnarray}
			\mathbf{E} &= E_0 & \sin(\omega t) \widehat{e_y}  \\
			\mathbf{B} &= \frac{E_0}{c}& \sin(\omega t) \widehat{e_z}
		\end{eqnarray}
		
		Which yields a perturbation which we can factor into its spatial term, $V(\rvec)$, and its sinusoidal time dependence, $\sin(\omega t )$
		
		\begin{eqnarray}
			\ham{'} &=&  q y E_0 \sin(\omega t )  \nonumber \\
			\ham{'} &=&  V(\rvec) \sin(\omega t ) 
		\end{eqnarray}
		
		With:
		\begin{equation}
			V(\rvec) = q y E_0 = p(\rvec) E_0
		\end{equation}
		
		Where $p$ encodes the intrinsic dipole moment. 
		
		As desired, this time-dependent perturbation exhibits sinusoidal time-dependence.  Transitions induced by a sinusoidal time-dependent perturbation are well studied \cite{Griffiths}, but before fully solving the system, it is useful to qualitatively understand the system. To do so, we can find the transition probability between states to first-order using off-diagonal elements of the perturbation following Fermi's Golden Rule:
		
		\begin{equation}
			\label{eq:FermisGoldenRule}
			T_{2 \rightarrow 1} \propto \normsq{ \bra{2} \ham{'} \ket{1} } \equiv \normsq{ \ham{'_{21}} } =  \normsq{ \ham{'_{12}} }
		\end{equation}
		
		By the hermeticity of $\ham{'}$, we find that for a single particle, the $\ket{1} \rightarrow \ket{2}$ transition is just as likely as the $\ket{2} \rightarrow \ket{1}$ transition. The net transition of a system (a net absorption or emission of photons) is only determined by the initial population of each state, $N_1$ and $N_2$.  It is ultimately this fact that explains why population inversions are so crucial to amplification by stimulated emission; in solving the system fully, we only learn how much the signal is amplified, whereas energy level populations determine whether we observe net absorption or emission. But the strength of amplification is important, as it allows us to infer many details about the system, so we will continue to fully solve for the complete transition probability.

		Our perturbing Hamiltonian has already been factored into a form: $\ham{'} = V(\rvec) cos(\omega t)$, and we are focusing on 2 energy levels with an energy level gap of $\hbar \omega_0$.  If the perturbation is sufficiently small ($\normsq{ V_{12} } \ll \hbar^2 \left(\omega - \omega_0\right)^2$), then we can solve for the transition probability to first-order:
		
		\begin{eqnarray}
			P_{2 \rightarrow 1} &\approx& \frac{\normsq{ V_{21} }}{\hbar^2} \frac{ \sin^2\left[ \left(\omega_0 - \omega \right) \frac{t}{2} \right]}{\left(\omega_0 - \omega \right)^2}  \nonumber \\
			 &=& \frac{\normsq{ p_{21} E_0} }{\hbar^2}   \frac{ \sin^2\left[ \left(\omega_0 - \omega \right) \frac{t}{2} \right]}{\left(\omega_0 - \omega \right)^2} \nonumber \\
		\end{eqnarray}
		
		We can simplify the expression slightly by introduction the notation $\sinc(x) \equiv \frac{\sin(x)}{x}$. For monochromatic light, this leaves us with the transition probability:
		
		\begin{equation}
			\label{eq:TransProbMonoSinc}
				P_{2 \rightarrow 1} = \frac{\normsq{ p_{21} E_0} }{\hbar^2}  \sinc^2\left[ \left(\omega_0 - \omega \right) \frac{t}{2} \right] \frac{t^2}{4}	 
		\end{equation}
		
		Before going much further, we will extend from a linearly polarized plane wave, to an isotropic bath of radiation.  To correct for the distribution of polarization orientations, we simply take dipole matrix element in the direction of the polarization, $\widehat{e_n}$, averaged over all possible directions:
		
		\begin{eqnarray}
			\label{eq:PolarizationGeneralization}
			\normsq{p_{21}} = \normsq{  \bra{2} q y \ket{1} } &\rightarrow& \langle \normsq{ \bra{2} q\rvec \ket{1} \cdot \widehat{e_n} } \rangle_{\widehat{e_n}}  \\
			\label{eq:PolarizationGeneralization_special}
			\normsq{p_{21}} &\rightarrow& \frac{\normsq{\mathbf{p}_{21}}}{3}
		\end{eqnarray}
		
		(For a more detailed treatment, see Footnote \footnote{This derivation followed the standard stimulated emission conventions \cite{Griffiths} in assuming an isotropic distribution of photon polarizations. This assumption is rarely true for astrophysical masers, which typically have very pronounced anisotropies, with beams of radiation over very small solid angles; see Section \ref{sec:AstroAmp} below.  It is possible to make a better estimation, but we only find constant factors of order unity, which do not provide any more insight.  If you are interested in making these geometric corrections, the substitution is still given by Substitution \ref{eq:PolarizationGeneralization}, but for some anisotropic distribution of polarizations. For instance, if integrated over only 1 polar angle (i.e. a circularly polarized plane wave) we find $\normsq{p_{21}} \rightarrow \frac{\sin^2\theta}{2} \normsq{\mathbf{p}_{21}}$ where $\theta$ is the angle between the dipole moment matrix element, $\mathbf{p}_{21}$, and the direction of propagation.}; most sources will quote the probability for an isotropic distribution of polarizations, and we will follow those conventions, understanding that it is possible to make suitable corrections for the orientation of specific systems. Those corrections provide only small geometric corrections, which give no new physical insight.)

		Generalizing to a distribution of polychromatic light is not much more difficult, and will be useful to astrophysical systems where we cannot control the radiation being amplified. Recognizing that $E_0^2$ is proportional to the monochromatic field energy density, $u = \frac{\epsilon_0 E_0^2}{2}$, we can rewrite Equation \ref{eq:TransProbMonoSinc} as:
		
		\begin{equation}
			\label{eq:TransProbMonoSinc_u}
				P_{2 \rightarrow 1} = \frac{2 \normsq{ \mathbf{p}_{21} }u }{3\epsilon_0 \hbar^2}  \sinc^2\left[ \left(\omega_0 - \omega \right) \frac{t}{2} \right] \frac{t^2}{4}	 
		\end{equation}	
		
		And simply replacing the monochromatic energy density with a energy density per unit frequency, $u \rightarrow \rho(\omega) d\omega$, we now have a generalization for stimulated emission by polychromatic light:

		\begin{equation}
			\label{eq:P_12_dimless_integral}
			P_{2 \rightarrow 1} = \frac{2 \normsq{ \mathbf{p}_{21}} }{3\epsilon_0 \hbar^2} \int_0^\infty \rho(\omega) \sinc^2 \left[ \left(\omega_0 - \omega \right) \frac{t}{2} \right] \frac{t^2}{4} d\omega
		\end{equation}		
		
		Our work is not yet complete; we can do better than just leaving ourselves with an integral over an unknown distribution $\rho(\omega)$.  Noticing that the $\sinc^2$ gain profile will be become more sharply peaked at $\omega_0$ as time passes, we can assume that after a long time has passed, $\sinc^2$ will be sensitive to $\rho(\omega)$ only at $\omega = \omega_0$ (the $\sinc^2$ term will effectively have become proportional to a delta function in frequency-space). This allows us to pull $\rho(\omega_0)$ out of the integral:
		
		\begin{equation}
			\label{eq:Prob_function_t}
			P_{2 \rightarrow 1} = \frac{2 \normsq{ \mathbf{p}_{21}} }{3\epsilon_0 \hbar^2} \rho(\omega_0) \int_0^\infty \sinc^2 \left[ \left(\omega_0 - \omega \right) \frac{t}{2} \right] \frac{t^2}{4} d\omega
		\end{equation}
		
		We can now non-dimensionalize it, using $\mu = (\omega - \omega_0) \frac{t}{2}$:
		
		\begin{eqnarray}
			P_{2 \rightarrow 1} &=& \frac {2 \normsq{ \mathbf{p}_{21}} }{3 \epsilon_0 \hbar^2} \rho(\omega_0) \frac{t}{2} \int_0^\infty  \sinc^2 \left[ \left(\omega_0 - \omega \right) \frac{t}{2} \right] d\left(\frac{\omega t}{2}\right) \nonumber \\
			 &=& \frac{\normsq{ \mathbf{p}_{21}} }{3\epsilon_0 \hbar^2} \rho(\omega_0)t \int_{-\omega_0 t/2}^\infty  \sinc^2 \left[ \mu \right] d\mu
		\end{eqnarray}
		
		But we are still left with a definite integral dependent on physical conditions (namely the energy gap, $\hbar \omega_0$), and we want a closed-form algebraic expression for the generic case. Note the limits of integration are far from the center of the $\sinc^2$ profile, and after a long time has passed contributions far from the center will be negligibility small. This safely allows us to extend the limits of integration to $\omega \in (-\infty, \infty)$:
		
		\begin{eqnarray}
			\label{eq:Prob_function_t}
			P_{2 \rightarrow 1} &=& \frac{ \normsq{ \mathbf{p}_{21}} }{3\epsilon_0 \hbar^2} \rho(\omega_0) t \int_{-\infty}^\infty  \sinc^2 \left[ \mu \right] d\mu \nonumber \\
			&=& \frac{ \normsq{ \mathbf{p}_{21}} }{3 \epsilon_0 \hbar^2} \rho(\omega_0) t \pi
		\end{eqnarray}
		
		One concern at this point is that we waited an arbitrarily long time in order to allow the $\sinc^2$ profile to become sharply peaked around $\omega_0$, but now we see that probability grows linearly with time. This seems to suggest that the probability is unbounded, allowing a transition probability greater than 1 if we only waited long enough.  This is just an artifact of the first-order approximation.  For a truly small perturbation, we require:
		
		\begin{equation}
			\label{eq:SmallCondition}
			\frac{\normsq{ \mathbf{p}_{21} } }{\epsilon_0 \hbar^2} \rho(\omega_0) t \ll 1
		\end{equation} 
		
		(This connects to the monochromatic perturbation requirement: $\normsq{ V_{12} } \ll \hbar^2 \left(\omega - \omega_0\right)^2$ but the full derivation is inconsequential to astrophysical masers. In future sections we will not be interested the evolution of a single state over a long period of time, but rather the average lifetime of an ensemble of states)
		
		One benefit of the linear growth of $P_{1 \rightarrow 2}$ is that it suggest a probabilistic \emph{transition rate}:
		
		\begin{equation}
			\label{eq:Rate12}
			R_{2\rightarrow 1} \equiv \frac{\partial P(t)}{\partial t} = \frac{\pi  \normsq{ \mathbf{p}_{21}}}{3\epsilon_0 \hbar^2} \rho(\omega_0 )
		\end{equation}
				
		Two things should be noticed about Equation \ref{eq:Rate12}:
		\begin{enumerate}
			\item $R_{1\rightarrow 2} = R_{2\rightarrow 1}$, as anticipated by Equation \ref{eq:FermisGoldenRule}.  This means stimulated emission of a single excited particle is equally as likely as absorption by a single unexcited particle.
			\item This is a first-order approximation for small perturbations (expansion criterion given by Equation \ref{eq:SmallCondition}). While initially the intensity of light, $\rho(\omega_0)$, might be small, it will amplify the very frequency to which it is most sensitive.  In Section \ref{sec:AstroAmp}, we'll see exactly how this can lead to amplification, and how amplification can lead to a regime of saturation, in which $\rho(\omega_0)$ is no longer ``small.''
			\item $R_{1\rightarrow 2}$ appears to be sensitive to $\rho(\omega)$ only at resonance, $\omega = \omega_0$.  The derivation thus far predicts only delta-function sensitivity in the maser and a corresponding delta-function emission profile. Section \ref{sec:LineNarrowing} will discuss the non-zero widths in the sensitivity and absorption profiles.
		\end{enumerate}
		
		\subsubsection{Amplification}
			\label{sec:AstroAmp}
		Our work does not stop with the single particle transition rate (Eq. \ref{eq:Rate12}). We want to understand the behavior of an ensemble of states $\ket{1}$ and $\ket{2}$, rather than the evolution of a single particle.  Under constant irradiation, Equation \ref{eq:Rate12} predicts a single particle will transition back-and-forth between energy levels. We want steady-state amplification, which we only observe with population inversions. For this section, we will adopt the traditional Einstein notation for stimulated emission and follow the work of Goldreich and Kwan \cite{GoldreichKwan} in solving for how masers amplify signals over distances.
		
		First it is beneficial to define some notation. We will first switch to using variables and units more familiar to astronomers. For frequency we will use $\nu = \frac{\omega}{2\pi}$, and instead of energy density per unit frequency, $\rho(\omega)$, we use specific intensity, $I_\nu$ (specific intensity and energy density are related by: $\rho(\nu) = \frac{1}{c} \int I_\nu(\hat{\Omega}) d\Omega$). In the particular application of masers, the radiation will be a highly beamed across a small solid angle --- we will approximate it as a beam of constant intensity over a solid angle, $\Delta\Omega$.  This leaves a mean intensity $J_\nu = I_\nu \frac{\Delta \Omega}{4 \pi}$. (A more precise treatment of beam shape is possible, but does not provide any new insight.)
		
		Our ensemble of particles will be built entirely of states in 2 energy levels, $E_1$ and $E_2$ such that $E_1 < E_2$.  These states have corresponding number densities, $n_1$, $n_2$.
		
		There is a chance that an excited state, $\ket{2}$, will spontaneously decay (in the absence of external radiation) to the lower energy state, $\ket{1}$, releasing a photon in the process. This rate is proportional to the number density of particles in state $\ket{2}$, with a proportionality constant of $A_{21}$. The rate of spontaneous decay is therefore: $A_{21} n_2$.
		
		There is also the possibility of stimulated transitions. These are the transitions derived in Section \ref{sec:derivation_transition_rates}, culminating with the rate equation (Eq. \ref{eq:Rate12}).  Whereas $R_{2 \rightarrow 1}$ contained both intrinsic information on the quantum states $\ket{1}$ and $\ket{2}$ as well as information about the external radiation, $\rho(\omega_0)$, we can factor out the intrinsic properties ($B_{21}$) from the external conditions ($J_\nu$), so that transitions occur at rates of $B_{12} n_1 J_{\nu_0}$ and $B_{21} n_2 J_{\nu_0}$ for absorption and stimulated emission respectively. Explicitly, this factorization is given by:
		
		\begin{equation}
			\label{eq:B21}
			B_{21} = \frac{R_{2\rightarrow 1}}{\rho(\omega_0)} = \frac{\pi    \normsq{\mathbf{p}_{21}}}{3 \epsilon_0 \hbar^2}
		\end{equation}
		
		Just as $R_{2 \rightarrow 1} = R_{1 \rightarrow 2}$, we now find $B_{12} = B_{21}$.
		
		In the literature of quantum absorption and emission, $A_{21}$ and $B_{21}$ are commonly known as the \emph{Einstein coefficients} \cite{Griffiths}.   From Equation \ref{eq:B21} we know the form of $B_{21}$; knowing that they are Einstein coefficients then implies \cite{Griffiths}:
		
		\begin{equation}
			\label{eq:A21}
			A_{21} = \frac{8 \pi h \nu^3}{c^3} B_{21}
		\end{equation}
		
		While $B_{21}$ corresponds to transitions stimulated by external radiation, $A_{21}$ is independent of external radiation.  $A_{21}$, the spontaneous emission coefficient, can be interpreted as transitions stimulated by QED background fluctuations \cite{Griffiths}, which have an effective intensity: $\frac{8 \pi h \nu^3}{c^3}$. At low frequencies (such as microwave frequencies) the factor of $\nu^3$ will suppress spontaneous emission, and we will neglect it in our calculations. 
		
		So far we have treated the system as a closed system (we have only considered transitions directly between $\ket{1}$ and $\ket{2}$).  In order to provide steady-state population inversions we need to generalize to an open system, one which allows particles to enter and leave the system. We define  $\lambda_i$, the \emph{pumping rate}, for the rate of addition of particles of state $\ket{i}$.  Likewise, $\Gamma n_i$, the \emph{quenching rate}, will describe the removal of particles of state $\ket{i}$.
		
		(For conditions common to astrophysical masers, $\Gamma \gg A_{21}$ \cite{GoldreichKwan}.  Any excited state which does not undergo stimulated emission will most likely be quenched long before it undergoes spontaneous emission. This is another indicator that we can safely neglect $A_{21}$. Whenever an equation is introduced in the following derivation, we will first show how to properly include $A_{21}$ and then drop it from the subsequent algebra.)
		
		We now have our model for how the populations of our two state system evolves:
		
		\begin{eqnarray}
			\label{eq:ParticleConservation1}
			\frac{d n_1}{dt} &=& \lambda_1 - \Gamma n_1 
			 - B_{21} J_{\nu_0} \left(n_1 - n_2 \right) + A_{21}n_2\\
			\label{eq:ParticleConservation2} \frac{d n_2}{dt} &=& \lambda_2 - \Gamma n_2  
			 - B_{21} J_{\nu_0} \left(n_2 - n_1 \right) - A_{21}n_2 
		\end{eqnarray}
		
		We are interested with the transitions, only insofar as they give us information about amplification of the stimulating radiation. For that, we build an equation describing the radiative transport for plane-wave radiation traveling in $\widehat{e_x}$:

		\begin{equation}
			\label{eq:RadiativeTransport}
			\frac{dI_{\nu_0}(x)}{dx} = \frac{h c}{4 \pi} \left( \left[n_2 - n_1\right] B_{21}I_{\nu_0}(x) + A_{21} n_2 \right)
		\end{equation}
		
		For a steady state amplification ($\frac{dn_i}{dt} = 0$) we can combine Equations \ref{eq:ParticleConservation1}, \ref{eq:ParticleConservation2},  and \ref{eq:RadiativeTransport} (dropping $A_{21}$):
		
		\begin{equation}
			\label{eq:Combined_Conservation_Transport}
			\frac{dI_{\nu_0}(x)}{dx} = \frac{h c}{4 \pi} \frac{\left[ \lambda_2 - \lambda_1 \right] }{ \left[ \Gamma + 2 B_{21} J_{\nu_0}(x) \right]} B_{21} I_{\nu_0}(x)
		\end{equation}
		
		Equation \ref{eq:Combined_Conservation_Transport} contains all the information we need about the differential amplification, but it is rather unwieldy in its current form.  In order to simplify Equation \ref{eq:Combined_Conservation_Transport}, we define an apparent opacity, $\kappa_{\nu_0}$, and saturation intensity, $I_s$ \cite{GoldreichKwan, Litvak}:
		
		\begin{eqnarray}
			\label{eq:kappa}
			\kappa_{\nu_0} &\equiv& \frac{h c}{4 \pi} \frac{\left[ \lambda_2 - \lambda_1 \right]  B_{21}}{\Gamma} \\
			\label{eq:Is}
			I_s &\equiv& \frac{2 \pi \Gamma}{\Delta \Omega B_{21}} 
		\end{eqnarray}
		
		Definitions  \ref{eq:kappa} and \ref{eq:Is} significantly simplify Equation \ref{eq:Combined_Conservation_Transport} to the form:
		
		\begin{equation}
			\label{eq:Combined_Conservation_Transport_simple}
			\frac{dI_{\nu_0}}{dx} = \frac{\kappa_{\nu_0} I_{\nu_0}}{1 + I_{\nu_0}/I_s}
		\end{equation}
		
		Which can be integrated to find:
		
		\begin{equation}
			\label{eq:IntensityGrowth_Full}
			I_{\nu_0}(x) = I_{\nu_0}(0) \exp\left[ \kappa_{\nu_0} x - \frac{I_{\nu_0}(x) - I_{\nu_0}(0)}{I_s} \right]
		\end{equation}
		
		\begin{figure}
			\centering
			\includegraphics[width= .9\linewidth]{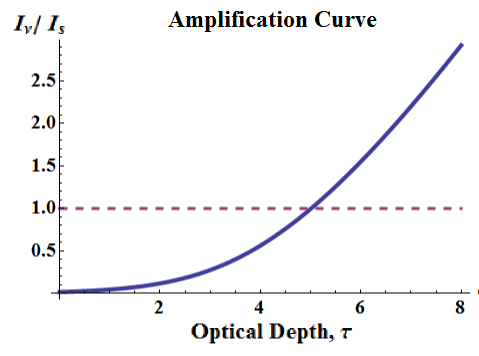}
			\caption{Optical depth, $\tau$, follows the convention $\tau = \kappa x$.  Note the exponential growth until $\tau \approx 5$ at which point it begins to level off to a linear growth for $I_{\nu_0} / I_s$. }
			\label{fig:AmplificationProfile}
		\end{figure}

		Equation \ref{eq:IntensityGrowth_Full} has two asymptotic behaviors (shown in Figure \ref{fig:AmplificationProfile}):
		\begin{enumerate}
		\item When the intensity is far below saturation of the medium ($I_\nu \ll I_s$) we observe exponential amplification (as a function of distance) of the original signal.  This regime will be the case more important to us in the coming sections because small changes in path length or opacity, $\kappa$, will produce significant changes in observed signals. 
		\item Close to, and above saturation ($I_{\nu_0} \gtrsim I_s$), the amplification slows down, growing linearly with distance.  At this point the intensity becomes so strong that every particle which enters the system will undergo a stimulated transition (now both quenching and spontaneous emission are negligible).  Taking both stimulated emission and absorption into account, we find a net emission rate of $\lambda_2 - \lambda_1$ which has decoupled from the intensity.  This shows that the saturation intensity, $I_s$,  is not the maximum intensity, but rather the point at which decoupling from intensity occurs.
		\end{enumerate}
		
		(If $\lambda_1 > \lambda_2$, then $\kappa_\nu < 0$, indicating absorptive conditions.  This also is equivalent to our statement that population inversions are required for net absorption.  In our current model of pumping and quenching, if $\lambda_1 > \lambda_2$, then we can never observe a population inversion. If $\lambda_1 < \lambda_2$ then we must observe a population inversion.  For the net absorptive regime, Equation \ref{eq:IntensityGrowth_Full} reproduces the Lambert-Beer Law: $I_\nu(x) = I_\nu e^{- \lvert \kappa_\nu \rvert x }$, which describes exponential absorption at thermodynamic equilibrium.)
		
		We have now found that under the conditions of a population inversion, there can be exponential amplification of faint signals.  This exponential amplification is sensitive to the total path length, the pumping rates ($\lambda_i$), the quenching rate ($\Gamma$), and the intrinsic electric dipole moment matrix element ($\mathbf{p}_{12}$; implicit in $B_{21}$, Eq. \ref{eq:B21}). Those sensitivities can allow us to trace small changes in amplifying systems, which we can use to probe the details of astrophysical systems.
		
		\subsection{Line Narrowing}
			\label{sec:LineNarrowing}
			In Section \ref{sec:derivation_transition_rates} we approximated the sensitivity and emission profiles of the transitions to be delta functions, centered at the energy of the transition between states. In Section \ref{sec:AstroAmp}, we saw amplification precisely at $I_{\nu_0}$ and nowhere else. In reality, we expect both the sensitivity and the emission to have some non-zero width. In this section we will take into account those non-zero widths. We will see \emph{line narrowing} at low saturations, and \emph{rebroadening} of spectral lines at large saturations.  Experimentally, line narrowing and subsequent rebroadening is an important feature --- it can help us identify amplifying systems due to unexpectedly narrow spectral lines. If not accounted for, one would infer unphysical low surface temperatures of emitting bodies due to the narrow line width.
			
			The non-zero width of the amplification can be explained by the velocities of the gas particles.  The thermal motion of gas particles will cause both stimulating and emitted radiation to be either red- or blue-shifted as viewed by different observers.  We will assume a thermal Boltzmann distribution of velocities (standard for a gas), with a velocity dispersion $\delta v = \sqrt{\frac{2 k_B T}{m}}$:
			
			\begin{equation}
				v = v_0 \exp\left[- \frac{v^2}{\delta v^2} \right]
			\end{equation}
			
			(We assume there is still a pumping mechanism pushing the energy level populations away from equilibrium, but that the pumping mechanism supplies particles with a thermal distribution of velocities.)
			
			If each particle was emitting at the resonant frequency, $\nu_0$ in its rest frame, we would expect the spectrum to be Doppler-broadened when viewed in the center-of-mass frame. This Doppler-broadening comes from redshifts at low velocites: $\frac{v}{c} = \frac{\nu - \nu_0}{\nu_0}$. Using that conversion we can find the spectrum of photon frequencies seen by the gas particles, and a characteristic Doppler-broadened width: 
			
			\begin{eqnarray}
				\label{eq:broadened_spectrum}
				I_\nu &=& I_0 \exp\left[- \frac{(\nu - \nu_0)^2}{\delta \nu_D^2} \right] \\
				\delta \nu_D &=& \nu_0 \frac{\delta v}{c}
			\end{eqnarray}
			
			Since amplification by stimulated emission is sensitive to precisely the same frequencies which are emitted, we can infer that the opacities, $\kappa_\nu$, and optical depths $\tau_\nu = \kappa_\nu x$ follow the same profile as Equation \ref{eq:broadened_spectrum}:
			
			\begin{eqnarray}
				\label{eq:broadened_opacity}
				\kappa_\nu &=& \kappa_0 \exp\left[- \frac{(\nu - \nu_0)^2}{\delta \nu_D^2} \right] \\
				\label{eq:broadened_depth}
				\tau_\nu = \kappa_\nu x &=& \tau_0 \exp\left[- \frac{(\nu - \nu_0)^2}{\delta \nu_D^2} \right]
			\end{eqnarray}
			
			This $\tau_\nu$ gives the extent of exponential amplification as a function of frequency, but we don't yet have a spectrum to amplify.  A reasonable assumption for astrophysical masers is that any preexisting spectral profiles were caused by Doppler-broadened spontaneous emission with a width of $\delta \nu_0$ \cite{Litvak1973, GoldreichKeeley}:
			
			\begin{equation}
				\label{eq:GaussianSpectrum}
				I_\nu(0) = I_0 \exp\left[ -\frac{ (\nu - \nu_0)  ^2 }{ \delta \nu_0^2} \right]
			\end{equation}
			
			Which at low saturations is amplified to:
			
			\begin{eqnarray}
				I_\nu &=& I_0 \exp \left[   -\frac{ (\nu-\nu_0)  ^2 }{ \delta \nu_0^2} +  \tau_0 \exp \left[ - \left( \frac{ \nu - \nu_0 }{\delta  \nu_D} \right) ^2 \right] \right] \nonumber \\ 
				&\approx& I_0 \exp\left[ \tau_0 \right] \exp \left[ -\left( \frac{\nu - \nu_0}{\delta \nu_0}\right)^2 \left( \frac{\delta \nu_0^2 }{\delta \nu_D^2} + \tau_0 \right) \right]
			\end{eqnarray}

			Which gives us an effective line width:
			
			\begin{equation}
				\label{eq:LineWidth}
				\delta \nu = \frac{\delta \nu_0}{\left( \frac{\delta \nu_0^2 }{\delta \nu_D^2} + \tau_0 \right)^{1/2}}
			\end{equation}
			
\begin{figure}
				\centering
				\includegraphics[width=.9\linewidth]{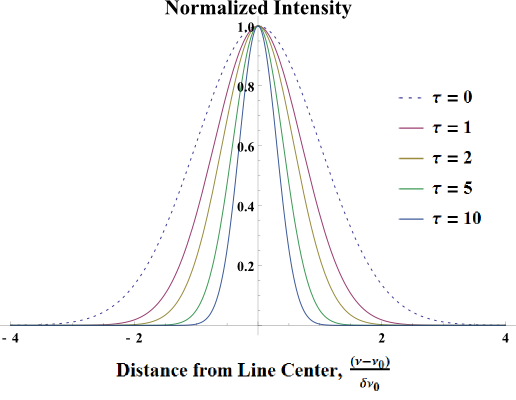}
				\caption{Simulated line narrowing of a Gaussian Dopper-broadened spectrum at low saturation.}
				\label{fig:GaussianNarrowing}
			\end{figure}

			Even if the amplifying gas has the same temperature (and thus the same velocity dispersion) as the radiation source gas (so $\frac{\delta \nu_0 }{\delta \nu_D} = 1$) we would still observe line narrowing! This line narrowing, of the form $\text{FWHM} \sim \delta \nu \propto \frac{1}{\left( 1 + \tau_0 \right)^{1/2}}$, can be seen in Figure \ref{fig:GaussianNarrowing}.  Such line narrowing has been clearly seen narrowing spectral features by at least a factor of 2-3 \cite{ElitzurBook}.  Without a population inversion we cannot explain one gas of thermally distributed velocities narrowing the radiation given off by another cloud of gas with the exact same thermal velocity dispersion.
			
			As masers begin to saturate though, the intensity at resonant frequency begins to grow linearly with depth, while the wings continue to grow exponentially. At this point line narrowing ceases, and the wings begin to grow relative to the peak, resulting in rebroadening of the spectral line. 
			
			It can be difficult to know whether a spectral line was simply initially narrow or has been narrowed through amplification by stimulated emission.  Looking at the widths of multiple lines can help identify lines which have been narrowed while the rest have remained untouched.  Another approach is using the observed width to infer a surface temperature of the radiation, and compare that to temperatures predicted by other measurements \cite{ElitzurBook}.

			A more complete treatment of line narrowing requires more complete models of opacity, source radiation spectrum,  velocity dispersion and geometry of the medium \cite{MaserThesis, Lekht, ElitzurBook}.  More complicate spectra begin to lose their Gaussian shape and even the symmetry in their tails.  Such effects cannot be explained simply by differences in the source emission. This further enables us to identify stimulated amplification; by comparing relative line widths, heights, and shapes of spectra we can partially discriminate between physical processes at the source and physical processes in the amplifying medium \cite{Weisberg01072005, Lekht}.
		
\section{Research Applications and Conclusion}
		\label{sec:ResearchApplications}
		In Section \ref{sec:AstroAmp} we found that small variations of the local conditions along the path of light could cause radiation to be very path-dependent.  This phenomenon is called \emph{anisotropic beaming}.  By itself, it is difficult to distinguish between anisotropic beaming of amplified radiation and radiation which was naturally collimated by another process.  Amplification by stimulated emission also has a special property of \emph{line narrowing}: amplified spectral lines have a much narrower width than their initial Doppler-broadened line shape.
		
		Anisotropic beaming and line narrowing allow us to recognize when amplification by stimulated emission is taking place.  The extent of these effects gives information about the physical environment.	This allows us a valuable look into poorly understood systems. 
		
		One particular application is the magnetic field configuration.  These magnetic fields, though Zeeman, paramagnetic and other effects, can induce energy level splitting while also making the amplifying system sensitive to particular polarizations of stimulating radiation \cite{Vlemmings}. The study of magnetic field lines and polarization in masers has been applied to both early- and late-type stellar objects, offering a valuable look into objects which are less-understood than more typical Main Sequence stars, such as the Sun \cite{ElitzurWeb}. Objects being studied include regions of rapid star formation, planetary nebulae, stellar envelops of asymptotic giant branch stars, supernovae remnants, and millisecond pulsars  \cite{PerezALMA, Vlemmings}.
		
		By understanding and studying stimulated microwave emissions, we gain new insight into physical environments, especially around young and old stars and compact stellar bodies.

\subsection*{Acknowledgments}
 
I would like to thank Ethan Stanley Dyer and Sophie Weber for their guidance and feedback during the course of writing this paper.  I would also like to thank Professor Thaler for his clarifications of technical aspects of understanding amplification by stimulated emission.

%\printbibliography{}

%\begin{thebibliography} {9}

%\end{thebibliography}


\begin{thebibliography}{15}
\expandafter\ifx\csname natexlab\endcsname\relax\def\natexlab#1{#1}\fi
\expandafter\ifx\csname bibnamefont\endcsname\relax
  \def\bibnamefont#1{#1}\fi
\expandafter\ifx\csname bibfnamefont\endcsname\relax
  \def\bibfnamefont#1{#1}\fi
\expandafter\ifx\csname citenamefont\endcsname\relax
  \def\citenamefont#1{#1}\fi
\expandafter\ifx\csname url\endcsname\relax
  \def\url#1{\texttt{#1}}\fi
\expandafter\ifx\csname urlprefix\endcsname\relax\def\urlprefix{URL }\fi
\providecommand{\bibinfo}[2]{#2}
\providecommand{\eprint}[2][]{\url{#2}}

\bibitem[{\citenamefont{Shklovsky}(1960)}]{Shklovsky}
\bibinfo{author}{\bibfnamefont{I.}~\bibnamefont{Shklovsky}},
  \emph{\bibinfo{title}{Cosmic Radio Waves}} (\bibinfo{publisher}{Harvard
  University Press}, \bibinfo{year}{1960}).

\bibitem[{\citenamefont{Elitzur}(2005)}]{ElitzurNat}
\bibinfo{author}{\bibfnamefont{M.}~\bibnamefont{Elitzur}},
  \textbf{\bibinfo{volume}{309}}, \bibinfo{pages}{71} (\bibinfo{year}{2005}),
  \urlprefix\url{http://www.sciencemag.org/content/309/5731/71.short}.

\bibitem[{\citenamefont{{Goldreich} and {Kwan}}(1974)}]{GoldreichKwan}
\bibinfo{author}{\bibfnamefont{P.}~\bibnamefont{{Goldreich}}} \bibnamefont{and}
  \bibinfo{author}{\bibfnamefont{J.}~\bibnamefont{{Kwan}}},
  \bibinfo{journal}{\apj} \textbf{\bibinfo{volume}{190}}, \bibinfo{pages}{27}
  (\bibinfo{year}{1974}).

\bibitem[{\citenamefont{Siegman}(1976)}]{Siegman}
\bibinfo{author}{\bibfnamefont{A.}~\bibnamefont{Siegman}},
  \emph{\bibinfo{title}{Lasers}} (\bibinfo{publisher}{University Science
  Books}, \bibinfo{year}{1976}), ISBN \bibinfo{isbn}{0-935702-11-03}.

\bibitem[{\citenamefont{Griffiths}(1995)}]{Griffiths}
\bibinfo{author}{\bibfnamefont{D.~J.} \bibnamefont{Griffiths}},
  \emph{\bibinfo{title}{Introduction to Quantum Mechanics}}
  (\bibinfo{publisher}{Prentice-Hall}, \bibinfo{year}{1995}), ISBN
  \bibinfo{isbn}{0-13-124405-1}.

\bibitem[{\citenamefont{{Litvak}}(1970)}]{Litvak}
\bibinfo{author}{\bibfnamefont{M.~M.} \bibnamefont{{Litvak}}},
  \bibinfo{journal}{\pra} \textbf{\bibinfo{volume}{2}}, \bibinfo{pages}{937}
  (\bibinfo{year}{1970}).

\bibitem[{\citenamefont{{Litvak}}(1973)}]{Litvak1973}
\bibinfo{author}{\bibfnamefont{M.~M.} \bibnamefont{{Litvak}}},
  \bibinfo{journal}{\apj} \textbf{\bibinfo{volume}{182}}, \bibinfo{pages}{711}
  (\bibinfo{year}{1973}).

\bibitem[{\citenamefont{{Goldreich} and {Keeley}}(1972)}]{GoldreichKeeley}
\bibinfo{author}{\bibfnamefont{P.}~\bibnamefont{{Goldreich}}} \bibnamefont{and}
  \bibinfo{author}{\bibfnamefont{D.~A.} \bibnamefont{{Keeley}}},
  \bibinfo{journal}{\apj} \textbf{\bibinfo{volume}{174}}, \bibinfo{pages}{517}
  (\bibinfo{year}{1972}).

\bibitem[{\citenamefont{Elitzur}(1992)}]{ElitzurBook}
\bibinfo{author}{\bibfnamefont{M.}~\bibnamefont{Elitzur}},
  \emph{\bibinfo{title}{Astronomical Masers}} (\bibinfo{publisher}{Springer},
  \bibinfo{year}{1992}), ISBN \bibinfo{isbn}{978-94-011-2394-5},
  \urlprefix\url{http://books.google.com/books?id=kajIH3t_750C&lpg=PA111&ots=d9_vRIbEm6&dq=line%20narrowing%20masers&pg=PA106#v=onepage&q=line%20narrowing%20masers&f=false}.

\bibitem[{\citenamefont{Archibald}(2006)}]{MaserThesis}
\bibinfo{author}{\bibfnamefont{A.~M.} \bibnamefont{Archibald}},
  \emph{\bibinfo{title}{Astrophysical masers}},
  \bibinfo{howpublished}{Radiative Processes in Astrophysics Term Paper}
  (\bibinfo{year}{2006}), \bibinfo{note}{{M}cGill University},
  \urlprefix\url{http://www.physics.mcgill.ca/~cumming/642/notes2006/term_papers/andrew_masers.pdf}.

\bibitem[{\citenamefont{Lekht et~al.}(2002)\citenamefont{Lekht, Silant’ev,
  Mendoza-Torres, and Tolmachev}}]{Lekht}
\bibinfo{author}{\bibfnamefont{E.}~\bibnamefont{Lekht}},
  \bibinfo{author}{\bibfnamefont{N.}~\bibnamefont{Silant’ev}},
  \bibinfo{author}{\bibfnamefont{J.}~\bibnamefont{Mendoza-Torres}},
  \bibnamefont{and}
  \bibinfo{author}{\bibfnamefont{A.}~\bibnamefont{Tolmachev}},
  \bibinfo{journal}{Astronomy Letters} \textbf{\bibinfo{volume}{28}},
  \bibinfo{pages}{89} (\bibinfo{year}{2002}), ISSN \bibinfo{issn}{1063-7737},
  \urlprefix\url{http://dx.doi.org/10.1134/1.1448845}.

\bibitem[{\citenamefont{Weisberg et~al.}(2005)\citenamefont{Weisberg, Johnston,
  Koribalski, and Stanimirović}}]{Weisberg01072005}
\bibinfo{author}{\bibfnamefont{J.~M.} \bibnamefont{Weisberg}},
  \bibinfo{author}{\bibfnamefont{S.}~\bibnamefont{Johnston}},
  \bibinfo{author}{\bibfnamefont{B.}~\bibnamefont{Koribalski}},
  \bibnamefont{and}
  \bibinfo{author}{\bibfnamefont{S.}~\bibnamefont{Stanimirović}},
  \textbf{\bibinfo{volume}{309}}, \bibinfo{pages}{106} (\bibinfo{year}{2005}),
  \urlprefix\url{http://www.sciencemag.org/content/309/5731/106.abstract}.

\bibitem[{\citenamefont{{Vlemmings}}(2007)}]{Vlemmings}
\bibinfo{author}{\bibfnamefont{W.~H.~T.} \bibnamefont{{Vlemmings}}}, in
  \emph{\bibinfo{booktitle}{IAU Symposium}}, edited by
  \bibinfo{editor}{\bibfnamefont{J.~M.} \bibnamefont{{Chapman}}}
  \bibnamefont{and} \bibinfo{editor}{\bibfnamefont{W.~A.} \bibnamefont{{Baan}}}
  (\bibinfo{year}{2007}), vol. \bibinfo{volume}{242} of
  \emph{\bibinfo{series}{IAU Symposium}}, pp. \bibinfo{pages}{37--46},
  \eprint{0705.0885}.

\bibitem[{\citenamefont{Elitzur}()}]{ElitzurWeb}
\bibinfo{author}{\bibfnamefont{M.}~\bibnamefont{Elitzur}},
  \emph{\bibinfo{title}{Masers, interstellar and circumstellar, theory}},
  \urlprefix\url{http://ned.ipac.caltech.edu/level5/ESSAYS/Elitzur/elitzur.html}.

\bibitem[{\citenamefont{{P\'erez-S\'anchez, A. F.} and {Vlemmings, W. H.
  T.}}(2013)}]{PerezALMA}
\bibinfo{author}{\bibnamefont{{P\'erez-S\'anchez, A. F.}}} \bibnamefont{and}
  \bibinfo{author}{\bibnamefont{{Vlemmings, W. H. T.}}},
  \bibinfo{journal}{A\&A} \textbf{\bibinfo{volume}{551}}, \bibinfo{pages}{A15}
  (\bibinfo{year}{2013}),
  \urlprefix\url{http://dx.doi.org/10.1051/0004-6361/201220735}.

\end{thebibliography}
\end{document}